\documentclass[aps,pra,amsmath,amsfonts,amssymb,twocolumn,showpacs]{revtex4}

\usepackage{amssymb}
\usepackage{graphicx,psfrag,color}
\usepackage{dcolumn}
\usepackage{bm}
\usepackage{mathbbol}

\begin{document}
\flushbottom

\title{Fighting noise with noise in realistic quantum teleportation}

\author{Raphael Fortes}
\author{Gustavo Rigolin}
\email{rigolin@ufscar.br}
\affiliation{Departamento de F\'isica, Universidade Federal de
S\~ao Carlos, 13565-905, S\~ao Carlos, SP, Brazil}

\date{\today}

\begin{abstract}
We investigate how the efficiency of the quantum teleportation protocol is 
affected when the qubits involved in the protocol are subjected to noise or decoherence.
We study all types of noise usually encountered in real world implementations of 
quantum communication protocols, namely, the bit flip, phase flip (phase damping), depolarizing, 
and amplitude damping noise. Several realistic scenarios are studied in which a part or all of the 
qubits employed in the execution of the quantum teleportation protocol are subjected to the same 
or different types of noise. We find noise scenarios not yet known in which more noise or less entanglement
lead to more efficiency. Furthermore, we show that if noise is unavoidable it is better to subject 
the qubits to different noise channels in order to obtain an increase in the efficiency of the protocol.
\end{abstract}


\maketitle

\section{Introduction} 

In quantum mechanics a physical system is described by its wave function (quantum state),
whose knowledge allows us to make predictions about the odds of the results of any measurement
implemented on the system. Also, all physical properties associated with the system are in principle 
derivable from its wave function. The goal of the 
quantum teleportation protocol \cite{ben93,vai94,bra98,bow97,bos98,fur98} is to transfer  
the wave function describing a system in one location (Alice) to another system in a different place (Bob),
without knowing the wave function. At the end of the protocol Alice's system is no longer described
by its original wave function, which now describes Bob's system.   

A key ingredient in quantum teleportation is a quantum channel connecting Alice and Bob
that is supposed to be a maximally entangled bipartite pure state \cite{ben93,vai94,bra98,bow97,bos98,fur98}. 
However, in any realistic implementation of the protocol noise is inherently present and it affects the entangled state during its 
transmission to Alice and Bob. The main effect of noise is to turn pure states into mixed ones.
A standard solution to overcome the effects of noise is called entanglement distillation \cite{ben96}, where
several copies of non-maximally entangled mixed states are needed to ``distill'' a pure one via local operations and
classical communication (LOCC).

Another important strategy to overcome this limitation is based on the direct use of the 
quantum channel connecting Alice and Bob \cite{guo00,bow01,alb02,lee02,tak12,ban12,kno14}, 
without resorting to distillation techniques, in which modifications in the standard teleportation
protocol are made to maximize its performance.  The formal solution to the optimal protocol 
when the most general quantum operations are allowed is given in Ref. \cite{tak12}. However, it is not, in general, an easy experimental 
task to implement the optimal quantum operations, and more feasible options leading
to an enhancement in the performance of the standard teleportation protocol are desired \cite{hor00,ban02,yeo08,ban12,kno14}.

In this paper we present a strategy to improve the efficiency of the teleportation protocol when both the qubit 
to be teleported and the quantum channel are subjected to noise. Our strategy consists of minimalistic
modifications in the standard teleportation protocol and an additional ingredient, the cheapest one, namely, 
more noise. We want to take advantage of the inevitable fact that noise is always present in any real world 
implementation of quantum communication protocols. 
We want to beat the decrease in the efficiency of the protocol due to noise 
with noise. Indeed, we show for several
realistic situations that it is possible by adding more noise or by choosing the right noisy environment to increase 
substantially the performance of the quantum teleportation protocol when compared to the less noisy case.

In order to be as general as possible, we analyze the most common noise channels one usually encounters in
the laboratory as well as many scenarios in which one, two, or all three qubits employed in the teleportation
protocol are acted by noise. We show several cases in which more noise,
less entanglement, or different noise acting on different qubits can enhance the efficiency of the teleportation protocol 
(Sec. \ref{results}). We also show that different channels with initially the same amount of entanglement give different 
efficiencies when subjected to the same type of noise (Sec. \ref{difchannel}).
The physical ideas and mathematical formalism needed to understand those results are given in the 
first part of this paper, Secs. \ref{secI}, \ref{noise}, and \ref{fidelity},
where we present, respectively, the standard teleportation protocol in the density matrix formalism, list all noise channels
we use and explain how they affect the qubits of the protocol, and show how to compute its efficiency 
in an input-state-independent way.

\section{Quantum teleportation protocol}
\label{secI}

In what follows we present the standard teleportation protocol \cite{ben93} in the density matrix formalism.
This formalism allows us to easily implement the quantum operations representing the several kinds of noise that may
affect the qubits involved in the protocol. 

The qubit to be teleported from Alice to Bob, hereafter called input qubit, is given by 
$|\psi\rangle= a|0\rangle + b|1\rangle$, with $|a|^2+|b|^2=1$. Its density matrix is 
\begin{equation}
\rho_{in} = |\psi\rangle\langle\psi | =
\left(
\begin{array}{cc}
|a|^2 & ab^* \\
a^*b & |b|^2
\end{array}
\right),
\end{equation}
where the subscript $in$ means ``input'' and $*$ denotes complex conjugation.  
The quantum channel shared between Alice and Bob is 
$|B_1^\theta\rangle=\cos\theta|00\rangle+\sin\theta|11\rangle$ and its density matrix 
in the base $\{|00\rangle,|01\rangle,|10\rangle,|11\rangle\}$ is 
\begin{equation}
\rho_{ch} = |B_1^\theta\rangle\langle B_1^\theta | =
\left(
\begin{array}{cccc}
\cos^2\theta & 0 & 0 & \sin\theta\cos\theta \\
0& 0 & 0 & 0 \\
0& 0 & 0 & 0 \\
\sin\theta\cos\theta & 0 & 0 & \sin^2\theta
\end{array}
\right).
\label{channel}
\end{equation}
Here the subscript $ch$ means ``channel'', the first qubit is with Alice, and the second one with Bob.
Note that when $\theta = \pi/4$ we have the Bell state $|\Phi^+\rangle$, a maximally entangled state whose
entanglement can be quantified by the entanglement monotone called concurrence (C) \cite{woo98}. $C \in [0,1]$
and in the present case it is equal to $C=|\sin(2\theta)|$. Note that we do not set $\theta$ to a predetermined value since
it will be a free parameter that one can adjust to maximize the efficiency of the noisy teleportation
protocol.

In order to start describing the protocol we also need to define a set of four orthonormal states onto which 
Alice will project the qubits with her,
\begin{eqnarray}
|B_1^\varphi\rangle&=&\cos\varphi|00\rangle+\sin\varphi|11\rangle, \label{B1} \\
|B_2^\varphi\rangle&=&\sin\varphi|00\rangle-\cos\varphi|11\rangle, \\
|B_3^\varphi\rangle&=&\cos\varphi|01\rangle+\sin\varphi|10\rangle, \label{B3} \\
|B_4^\varphi\rangle&=&\sin\varphi|01\rangle-\cos\varphi|10\rangle.
\end{eqnarray}
When $\varphi=\pi/4$ we recover the four Bell states, namely, $|\Phi^+\rangle, 
|\Phi^-\rangle, |\Psi^+\rangle$, and $|\Psi^-\rangle$. Here we also let $\varphi$
be a free parameter that will be chosen in order to optimize the efficiency of the 
teleportation. 
Using these four states we can define the projectors associated with the 
projective measurements that Alice implements in the execution of the 
protocol \cite{note1},
\begin{equation}
P_j^\varphi=|B_j^\varphi\rangle\langle B_j^\varphi|, \hspace{.5cm} j=1,2,3,4.
\end{equation}

Let us now start describing the teleportation protocol \cite{ben93} in the density
matrix formalism. The initial total state
describing the three qubits before any quantum operation is implemented is given by
\begin{equation}
\rho = \rho_{in} \otimes \rho_{ch}.
\label{step1}
\end{equation}
The first step of the protocol consists of Alice making a projective measurement on her
two qubits, namely, the input state and her share of the entangled channel. These qubits
are projected onto the basis $\{|B_j^\varphi\rangle\}$. After this measurement the 
total state (\ref{step1}) changes to
\begin{equation}
\tilde{\rho}_j = \frac{P_j^\varphi \rho P_j^\varphi}{\mbox{Tr}[{P_j^\varphi \rho}]}, 
\label{nine}
\end{equation}
where $\mbox{Tr}$ means the trace operation.
The probability of occurrence of a particular $\tilde{\rho}_j$ is
\begin{equation}
Q_j = \mbox{Tr}[{P_j^\varphi \rho}].
\label{prob}
\end{equation}

In the next step of the protocol Alice informs Bob of which $|B_j^\varphi\rangle$ she measured. 
With this information Bob knows that his state is now given by
\begin{equation}
\tilde{\rho}_{_{B_j}} = \mbox{Tr}_{12}[\tilde{\rho}_j] = \frac{\mbox{Tr}_{12}[P_j^\varphi \rho P_j^\varphi]}{Q_j}, 
\end{equation}
in which $\mbox{Tr}_{12}$ means the partial trace on qubits $1$ and $2$, the ones with Alice. 

The protocol ends with Bob implementing a unitary operation $U_j$ on his state.
Thus, the final state with Bob is 
\begin{equation}
\rho_{_{B_j}}= U_j\tilde{\rho}_{_{B_j}}U_j^\dagger =  \frac{U_j\mbox{Tr}_{12}[P_j^\varphi \rho P_j^\varphi]U_j^\dagger}{Q_j}.
\label{twelve}
\end{equation}
The unitary correction that Bob must implement to finish the protocol depends not only on the measurement result 
of Alice but also on the quantum channel used in the protocol. For the present case, where $\rho_{ch}$ is given by Eq.~(\ref{channel}),
$U_1=\mathbb{1}$, with $\mathbb{1}$
the identity matrix, $U_2=\sigma_z, U_3=\sigma_x$, and $U_4=\sigma_z\sigma_x$, with $\sigma_z$ and $\sigma_x$ being the
standard Pauli matrices. 

It is worth mentioning that whenever $\theta=\varphi = \pi/4$ we recover the standard teleportation protocol \cite{ben93},
where $Q_j=1/4$ and $\rho_{_{B_j}}=\rho_{in}$ for any $j$. 
For different values of $\theta$ and $\varphi$, or when noise acts on the
quantum channel, $Q_j$ depends on the input 
state \cite{guo00} and an averaging over all possible input states is 
needed to estimate the efficiency of the protocol in an input-state-independent way
(see Sec. \ref{fidelity}).

\section{Modeling the noise}
\label{noise}

The interaction of a noisy environment with a qubit can be represented by a quantum operation
acting only on the Hilbert space associated with the qubit if we use the operator-sum representation formalism
\cite{kra83,nie00}. The operators $E_k$ representing a certain type of noise are usually called Kraus operators and for trace
preserving operations (conservation of probability) they satisfy the following condition,
\begin{equation}
\sum_{j=1}^nE_j^\dagger E_j = \mathbb{1},
\end{equation}
where for a qubit $1\leq n\leq 4$ and 
$\mathbb{1}$ is the identity matrix acting on the qubit's Hilbert space.  The action of the noise on the 
qubit $k$, described by the density matrix $\rho_k$, is
\begin{equation}
\rho_k \rightarrow \varrho_k = \sum_{j=1}^nE_j \rho_k E_j^\dagger. 
\label{actionOfnoise}
\end{equation}

In this work we will be dealing with four different types of noise that may act 
on a qubit, those we usually find in any realistic modeling of a noisy environment. 
We also assume that each qubit in the teleportation protocol is acted by noise in an independent way. 
A brief description
of the physical meaning of each kind of noise as well as its Kraus operators
are given below. More details can be found in Refs. \cite{kra83,nie00}.

\subsection{Bit flip}

The bit flip noise changes the state of a qubit from $|0\rangle$ to $|1\rangle$
or from $|1\rangle$ to $|0\rangle$ with probability $p$ and appears frequently in
the theory of quantum error correction. Its Kraus operators are
\begin{equation}
E_1 = \sqrt{1-p}\;\mathbb{1}, \hspace{.5cm} E_2 = \sqrt{p}\;\sigma_x.
\end{equation}

\subsection{Phase flip or phase damping}

The phase flip noise changes the phase of the qubit $|1\rangle$ to $-|1\rangle$ with 
probability $p$ and it is described by the following Kraus operators,
\begin{equation}
E_1 = \sqrt{1-p}\;\mathbb{1}, \hspace{.5cm} E_2 = \sqrt{p}\;\sigma_z.
\end{equation}
The phase flip operation is equivalent to the phase damping, a noise 
process considered genuinely quantum, describing the loss
of information in a quantum state without energy loss \cite{nie00}. It is the paradigmatic
modeling of decoherence, since the phase damping noise destroys the quantum superposition
of a given qubit, bringing to zero the off-diagonal elements of the density matrix.

\subsection{Depolarizing noise}

This important type of noise takes a qubit and replaces it with a completely mixed 
state $\mathbb{1}/2$ with probability $p$. It can be thought of as a
``white noise'', bringing any density matrix to the completely unpolarized 
state ($\langle \sigma_x \rangle=\langle \sigma_y \rangle=\langle \sigma_z \rangle=0$).
Its Kraus operators are
\begin{eqnarray}
E_1 =& \sqrt{1-3p/4}\;\mathbb{1}, & E_2 = \sqrt{p/4}\;\sigma_x, \\
E_3 =&\hspace{-.7cm} \sqrt{p/4}\;\sigma_y, & E_4 = \sqrt{p/4}\;\sigma_z.
\end{eqnarray}

\subsection{Amplitude damping}

The process of amplitude damping is important in modeling energy dissipation in 
several quantum systems and its Kraus operators are 
\begin{equation}
E_1 =
\left(
\begin{array}{cc}
1 & 0 \\
0 & \sqrt{1-p}
\end{array}
\right), 
\hspace{.5cm} 
E_2 = 
\left(
\begin{array}{cc}
0 & \sqrt{p} \\
0 & 0
\end{array}
\right).
\end{equation}
The quantity $p$ can be seen as the decay probability from the excited to the ground state 
for a two-level system. 
For certain specific decoherence models one gets
$p = 1 - e^{-t/T}$, with $t$ the time and $T$ the characteristic time of the decoherence
process.
 
\subsection{Noisy teleportation protocol}

The total density matrix describing the initial state, Eq.~(\ref{step1}), will change according to the types of noise that each qubit 
is independently subjected to. The density matrix after the action of the three sources of noise is obtained by successively
applying Eq.~(\ref{actionOfnoise}) to each one of the qubits,
\begin{widetext}
\begin{equation}
 \varrho = \sum_{i=1}^{n_{\!_I}} E_i(p_{\!_I})\left[\sum_{j=1}^{n_{\!_A}} F_j(p_{\!_A}) \left(\sum_{k=1}^{n_{\!_B}} G_k(p_{\!_B}) \rho\,  
 G_k^\dagger(p_{\!_B})\right)F_j^\dagger(p_{\!_A})\right] E_i^\dagger(p_{\!_I})
 = \sum_{i=1}^{n_{\!_I}}\sum_{j=1}^{n_{\!_A}}\sum_{k=1}^{n_{\!_B}}E_{ijk}(p_{\!_I},p_{\!_A},p_{\!_B})\rho E_{ijk}^\dagger(p_{\!_I},p_{\!_A},p_{\!_B}),
\label{noiseRho}
 \end{equation}
\end{widetext}
where $E_{ijk}(p_{\!_I},p_{\!_A},p_{\!_B})=E_i(p_{\!_I})\otimes F_j(p_{\!_A})\otimes G_k(p_{\!_B})$.
Here $E_i(p_{\!_I})=E_i(p_{\!_I})\otimes\mathbb{1}\otimes\mathbb{1}, F_j(p_{\!_A})=\mathbb{1}\otimes F_j(p_{\!_A})\otimes\mathbb{1}$,
and $G_k(p_{\!_B})=\mathbb{1}\otimes\mathbb{1} \otimes G_k(p_{\!_B})$ are, respectively, the Kraus operators associated with
the kind of noise acting on the input qubit, Alice's qubit of the entangled channel, and Bob's qubit of the entangled channel.
In the general case, different noises can act during different times (probabilities) and that is why we make explicit
the dependence of the Kraus operators on $p_{\!_I}, p_{\!_A}$, and $p_{\!_B}$. 
By using the density matrix $\varrho$, Eq.~(\ref{noiseRho}), in Eqs.~(\ref{nine}) to (\ref{twelve}) we get the relevant
quantities needed to analyze the quantum teleportation protocol 
when noise is present. And if we set $p_{\!_I}=p_{\!_A}=p_{\!_B}=0$ we recover the noiseless case.

\section{Efficiency of the noisy protocol}
\label{fidelity}

The figure of merit used here to quantify the efficiency of the protocol is the fidelity \cite{uhl76}.
Since the benchmark state (input state) is initially pure, the fidelity can be written
as
\begin{equation}
F_j = \mbox{Tr}[\rho_{in}\varrho_{_{B_j}}]=\langle \psi | \varrho_{_{B_j}} | \psi \rangle,
\end{equation}
where $\varrho_{_{B_j}}$ is given by Eq.~(\ref{twelve}) with $\rho$ replaced by the noisy state $\varrho$.
The fidelity ranges from zero to one and its maximal value occurs whenever the output ($\varrho_{_{B_j}}$) is equal (up to an irrelevant global phase)
to the input state ($| \psi \rangle$) and it is zero when the two states are orthogonal.

In order to take into account the fact that each state $\varrho_{_{B_j}}$ may occur with different probabilities,
we define the average fidelity as
\begin{equation}
\overline{F} = \sum_{j=1}^4 Q_jF_j,
\label{F1}
\end{equation}
with $Q_j$, Eq.~(\ref{prob}), being the probability of Bob getting the state $\varrho_{_{B_j}}$.

When noise is present, or when we have non-maximally entangled channels \cite{guo00}, 
$\overline{F}$ depends on the input state $\rho_{in}$. Therefore, in order to quantify the
efficiency of the protocol in a way that is independent of a particular input state, we have to assume a probability 
distribution $\mathcal{P}$ for the pool of input states available to Alice. Here we assume
a uniform distribution, i.e., any qubit is equally probable to be 
picked as an input state in the teleportation protocol. 

Being more specific, let us write without loss of generality an arbitrary qubit as
\begin{equation}
|\psi\rangle = |a||0\rangle + |b|e^{ic}|1\rangle,
\label{relative}
\end{equation}
where $|a|$ and $|b|$ are absolute values of $a$ and $b$, and $c$ is a real number
representing the relative phase between $a$ and $b$. State (\ref{relative}) is equivalent to
$a|0\rangle+b|1\rangle$ for they only differ by a global phase with no physical significance. 
Since $|a|^2+|b|^2=1$, we can choose
$|a|^2$ and $c$ as our independent variables and write $\mathcal{P}(|a|^2,c)$ for the chance of
obtaining a qubit with relative phase $c$ and having probability $|a|^2$ of being detected
in the state $|0\rangle$. Note that $c\in [0,2\pi]$ and $0\leq |a|^2\leq 1$. The normalization condition
for the probability density $\mathcal{P}(|a|^2,c)$ is
\begin{equation}
\int_{0}^{2\pi}\int_{0}^{1}\mathcal{P}(|a|^2,c)\mathrm{d}|a|^2\mathrm{d}c =1,
\end{equation}
and by assuming a uniform probability distribution, $\mathcal{P}(|a|^2,c)=\mbox{constant}$,
we immediately get
\begin{equation}
\mathcal{P}(|a|^2,c)=\frac{1}{2\pi}.
\label{uniform}
\end{equation}

Noting that the average value $\langle f \rangle$ of any function of $|a|^2$ and $c$ is 
$\int_{0}^{2\pi}\int_{0}^{1}f(|a|^2,c)\mathcal{P}(|a|^2,c)\mathrm{d}|a|^2\mathrm{d}c$, we
define 
\begin{equation}
\langle \overline{F} \rangle = \int_{0}^{2\pi}\int_{0}^{1}\overline{F}(|a|^2,c)\mathcal{P}(|a|^2,c)\mathrm{d}|a|^2\mathrm{d}c,
\label{F2}
\end{equation}
as the input state independent quantifier for the efficiency of the noisy teleportation protocol.
Here $\overline{F}(|a|^2,c)$ is given by Eq.~(\ref{F1}) and $\mathcal{P}(|a|^2,c)$ by
Eq.~(\ref{uniform}).

\section{Results}
\label{results}

\subsection{Noise in Alice's input qubit}

Assuming for the moment that the quantum channel is protected from noise ($p_{\!_A}=p_{\!_B}=0$) while 
the input state lies in a noisy environment ($p_{\!_I}\neq 0$) as described in Sec. \ref{noise}, the average
fidelity, Eq.~(\ref{F2}), for each type of noise can be written as
\begin{eqnarray}
\langle\overline{F}_{_{BF}}\rangle\hspace{-.1cm} &=&\hspace{-.1cm}\frac{2}{3} \left[1-\frac{p_{\!_I}}{2}+\frac{1-p_{\!_I}}{2}\sin (2\theta) \sin (2\varphi)\right]\hspace{-.1cm},
\label{BFZZ}\\ 
\langle\overline{F}_{_{PhF}}\rangle \hspace{-.1cm} &=&\hspace{-.1cm} \frac{2}{3} \left[1+\frac{1-2p_{\!_I}}{2} \sin(2\theta) \sin(2\varphi)\right]\hspace{-.1cm},
\label{PhFZZ}\\
\langle\overline{F}_{_{D}}\rangle \hspace{-.1cm} &=&\hspace{-.1cm} \frac{2}{3} \left[1-\frac{p_{\!_I}}{4}+\frac{1-p_{\!_I}}{2}\sin (2\theta) \sin (2\varphi)\right]\hspace{-.1cm},
\label{DZZ}\\
\langle\overline{F}_{_{AD}}\rangle \hspace{-.1cm} &=&\hspace{-.1cm} \frac{2}{3} \left[\hspace{-.05cm}1-\frac{p_{\!_I}}{4}+\frac{1}{2} \sqrt{1-p_{\!_I}} \sin (2\theta) \sin (2\varphi)\hspace{-.05cm}\right]\hspace{-.1cm}.
\label{ADZZ}
\end{eqnarray}
The subscripts in the left hand side of Eqs.~(\ref{BFZZ}) to (\ref{ADZZ}) represent the particular type of noise 
that the input state is subjected to, i.e., $BF \rightarrow$ Bit Flip,
$PhF \rightarrow$ Phase Flip, $D \rightarrow$ Depolarizing, and
$AD \rightarrow$ Amplitude Damping.

The first thing worth noticing analyzing Eqs.~(\ref{BFZZ}) to (\ref{ADZZ}) is that 
for all of them but $\langle\overline{F}_{_{PhF}}\rangle$ the optimal $\theta$ and $\varphi$
are such that $\theta=\varphi=\pm \pi/4$. This is true because $1-p_{\!_I}\geq 0$ and the maximum
is obtained if $\sin(2\theta) \sin(2\varphi)=1$. In the expression for $\langle\overline{F}_{_{PhF}}\rangle$,
however, $\sin(2\theta) \sin(2\varphi)$ is multiplied by $1-2p_{\!_I}$, which changes sign at $p_{\!_I}=1/2$.
Thus, for $p_{\!_I}<1/2$ the optimal settings are $\theta=\varphi=\pm \pi/4$ while 
for $p_{\!_I}>1/2$ we have $\theta=-\varphi=\pm \pi/4$. This means that if the input qubit is subjected 
to the phase flip noise for a considerable amount of time ($p_{\!_I}>1/2$), Alice can counterattack and improve
the efficiency of the teleportation protocol by changing the measuring basis ($\varphi \rightarrow -\varphi$).
Alternatively, Alice can keep using the original measuring basis and either she or Bob applies 
a $\sigma_z$ operation to the channel, changing it from $|B_1^\theta\rangle$ to
$|B_1^{-\theta}\rangle$.

Using the above optimal settings, Eqs.~(\ref{BFZZ}) to (\ref{ADZZ}) simplify to
\begin{eqnarray}
\langle\overline{F}_{_{BF}}\rangle &=&1-\frac{2p_{\!_I}}{3},
\label{BFZZOpt}\\ 
\langle\overline{F}_{_{PhF}}\rangle &=& \frac{2}{3} +\frac{|1-2p_{\!_I}|}{3},
\label{PhFZZOpt}\\
\langle\overline{F}_{_{D}}\rangle  &=& 1 - \frac{p_{\!_I}}{2},
\label{DZZOpt}\\
\langle\overline{F}_{_{AD}}\rangle &=& \frac{2}{3}-\frac{p_{\!_I}}{6}+\frac{1}{3} \sqrt{1-p_{\!_I}} .
\label{ADZZOpt}
\end{eqnarray}
\begin{figure}[!ht]
\includegraphics[width=8cm]{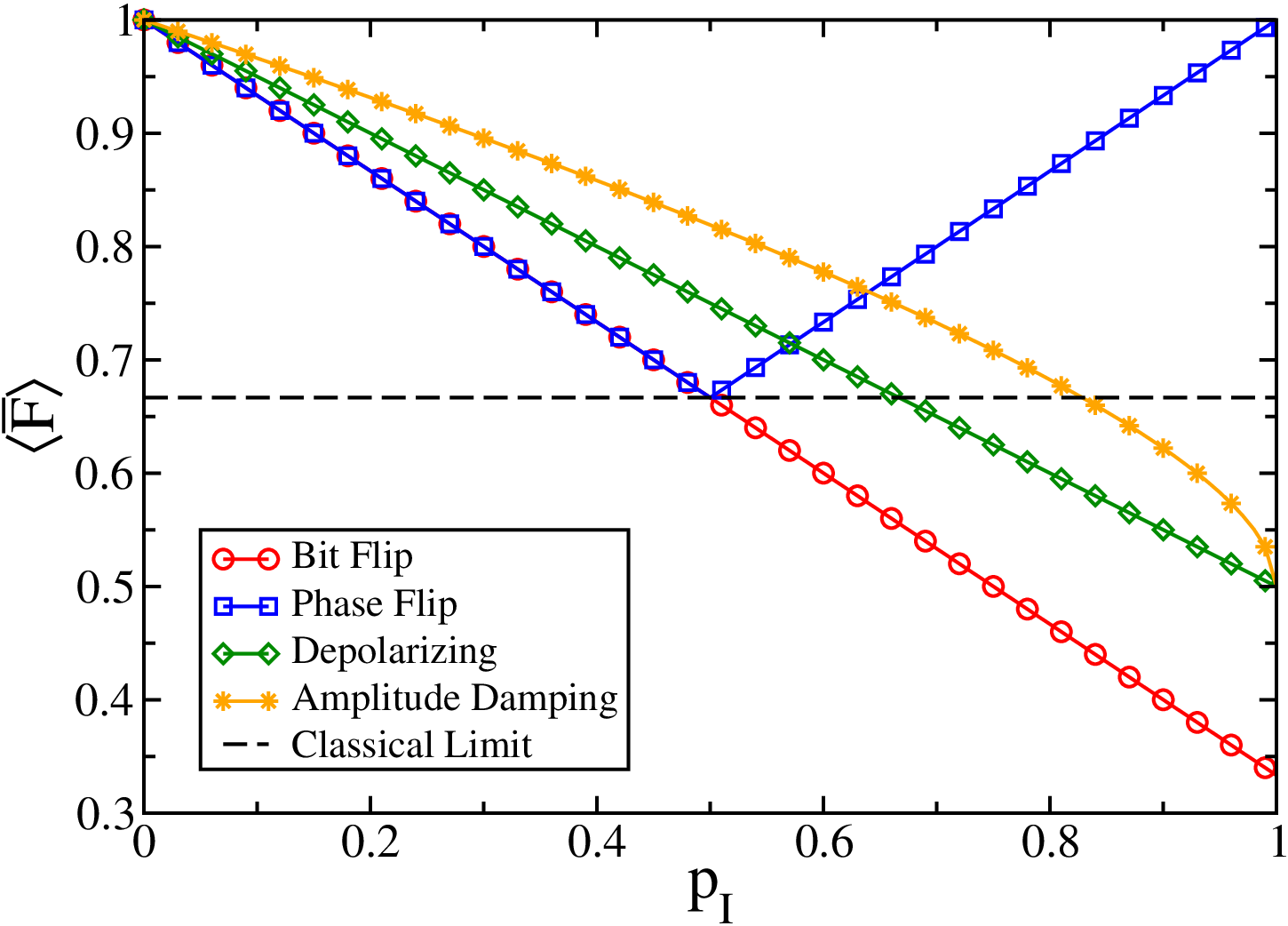}
\caption{\label{fig1}(color online) Efficiency of the teleportation protocol when only the 
input qubit is affected by a noisy environment, with $p_{\!_I}$ representing the probability for
the noise to act on the qubit. The dashed line, $\langle\overline{F}\rangle=2/3$, marks the value below which classical protocols
(no entanglement) give the same efficiency \cite{bra00}.}
\end{figure}

Looking at Fig.~\ref{fig1}, where we plot Eqs.~(\ref{BFZZOpt})-(\ref{ADZZOpt}), we see that
the bit flip noise is the most severe noise for all values of $p_I$ and that for $0\leq p_{\!_I} \leq 1/2$
the phase flip noise is as bad as the bit flip noise. On the other hand, from $p_{\!_I}=0$ to $p_{\!_I}\approx 0.6$, 
the amplitude damping is the least severe noise, followed by the depolarizing channel. For high values 
of $p_{\!_I}$, the phase flip noise gives the greatest average fidelity.

Let us now investigate what happens if one of the qubits belonging to the quantum channel is also subjected to noise. 
For definiteness we choose Bob's qubit, in addition to the input qubit, to lie in a noisy environment. However, 
it is not difficult to show that the same results follow if we choose Alice's qubit of the quantum channel. 
We also introduce the following notation in order to make it clear which qubits are subjected to noise. 
In the present case the optimal efficiency of the protocol, Eq.~(\ref{F2}), is written as
$\langle\overline{F}_{_{X,\varnothing,Y}}\rangle$, where the first subindex means that the input qubit is 
acted by noise $X$, the second one denotes that Alice's qubit of the quantum channel is not acted by noise, and
the third subindex means that Bob's qubit is acted by noise $Y$. Here $X$ and $Y$ can be any one of 
the four kinds of noise described previously. 

We start studying the case where the input qubit is always subjected to the bit flip noise while Bob's qubit
may lie in one of the four different types of noisy environments given in Sec. \ref{noise}. The optimal efficiencies
in those four cases are
\begin{eqnarray}
\langle\overline{F}_{_{BF,\varnothing,BF}}\rangle &=& 1-\frac{2}{3} (p_{\!_I}+p_{\!_B} - 2 p_{\!_I}p_{\!_B}),
\label{BFZBFOpt}\\ 
\langle\overline{F}_{_{BF,\varnothing,PhF}}\rangle&=& \frac{2}{3} -\frac{1}{3}[p_{\!_I}-(1-p_{\!_I})|1-2p_{\!_B}|],
\label{BFZPhFOpt}\\
\langle\overline{F}_{_{BF,\varnothing,D}}\rangle  &=& 1-\frac{p_{\!_B}}{2} - \frac{2}{3} p_{\!_I}(1 - p_{\!_B}),
\label{BFZDOpt}\\
\langle\overline{F}_{_{BF,\varnothing,AD}}\rangle &=& \frac{2}{3}-\frac{1}{3}\left[p_{\!_I}+\frac{p_{\!_B}}{2}(1-2p_{\!_I})(1-\cos (2\theta))\right.
\nonumber \\
&&\left. -(1-p_{\!_I})\sqrt{1-p_{\!_B}}\sin (2\theta)\right],
\label{BFZADOpt}
\end{eqnarray}
where the optimal parameters leading to Eqs.~(\ref{BFZBFOpt}) and (\ref{BFZDOpt}) are $\theta=\varphi=\pm\pi/4$ and to 
Eq.~(\ref{BFZPhFOpt}) are $\theta=\varphi=\pm\pi/4$ for $p_{\!_B}< 1/2$ and
$\theta=-\varphi=\pm\pi/4$ when $p_{\!_B}> 1/2$. In Eq.~(\ref{BFZADOpt}) we have $\varphi=\pi/4$ and the optimal 
$\theta$ given by the solution to $d\langle\overline{F}_{_{BF,\varnothing,AD}}\rangle/d\theta$ $=$ $0$, namely,
\begin{equation}
\tan(2\theta)=\frac{2(1-p_{\!_I})\sqrt{1-p_{\!_B}}}{p_{\!_B}(1-2p_{\!_I})},
\label{opttheta}
\end{equation}
such that $\sin(2\theta)>0$, $\cos(2\theta)>0$ for $p_{\!_I}<1/2$, and
$\cos(2\theta)<0$ for $p_{\!_I}>1/2$.

An interesting result worth mentioning here 
is that we have a scenario where \textit{less} entanglement means
\textit{more} efficiency. This happens when Bob's qubit is subjected to the amplitude damping noise (Eq.~(\ref{BFZADOpt})),
which was first noticed in Ref. \cite{ban12} when the input state is always pure. Here we show that even if the input state
is mixed we still have the same feature.
We can see that less entanglement means more efficiency looking at Eq.~(\ref{opttheta}), which 
tells us that the optimal $\theta$ is
$\pi/4$, the only solution meaning an initially maximally entangled channel, only when $p_{\!_B}=0$. 
For any other value of $p_{\!_B}$ the optimal $\theta$ is not $\pi/4$. Hence, whenever $p_{\!_B}\neq 0$, 
less entanglement leads to a better performance for the teleportation protocol in this case. 
 
\begin{figure}[!ht]
\includegraphics[width=8cm]{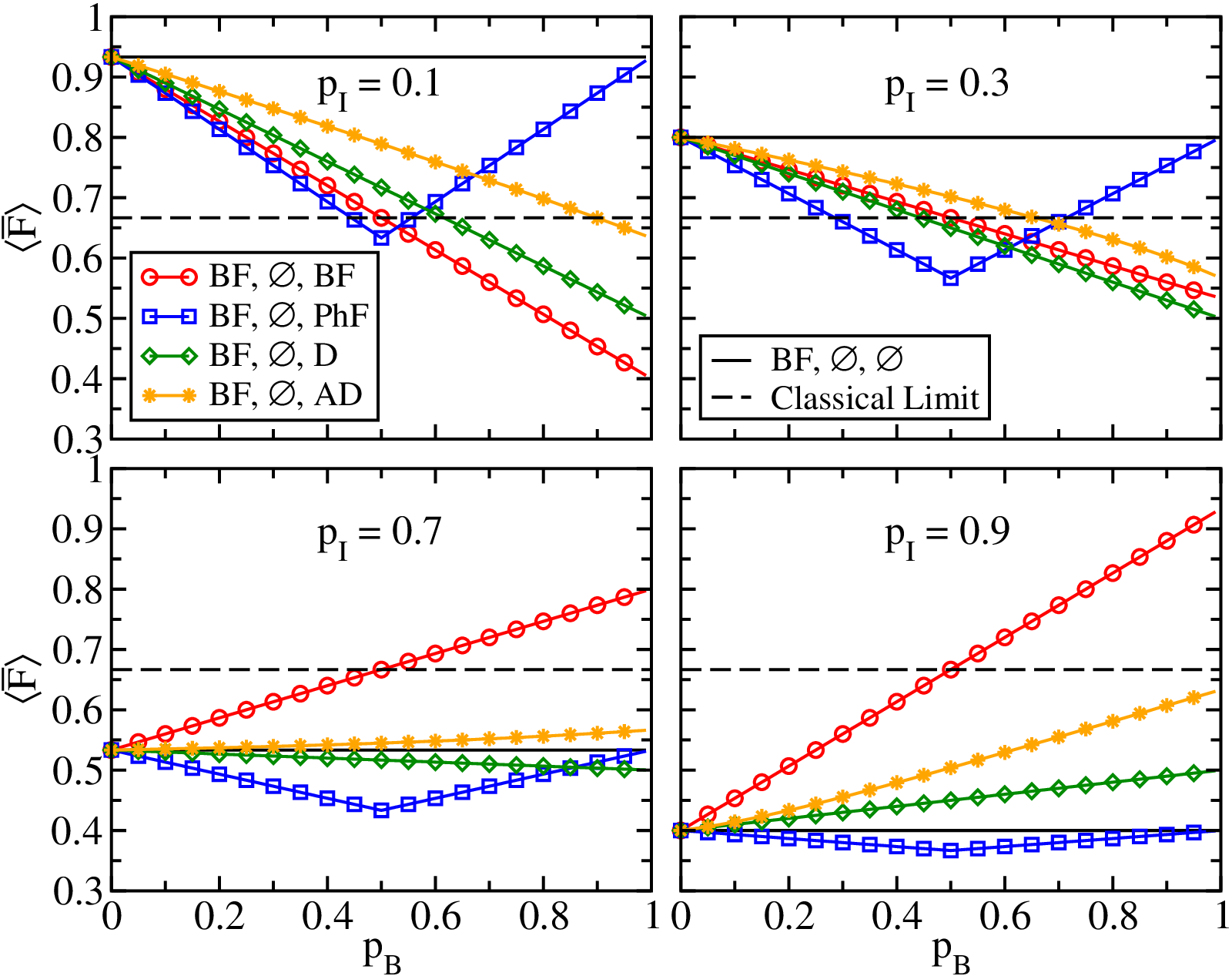}
\caption{\label{fig2}(color online) Efficiency of the teleportation protocol when both the  
input qubit ($p_{\!_I}$) and Bob's qubit ($p_{\!_B}$) are affected by a noisy environment.  
The dashed line, $\langle\overline{F}\rangle=2/3$, marks the value below which classical protocols
(no entanglement) give the same efficiency \cite{bra00}. Here the input qubit is always subjected
to the bit flip ($BF$) noise while Bob's qubit may suffer from several types of noise. }
\end{figure}

In Fig. \ref{fig2} we plot Eqs.~(\ref{BFZBFOpt}) to (\ref{BFZADOpt}) as a function of $p_{\!_B}$
for several values of $p_{\!_I}$. Looking at the two bottom panels of Fig. \ref{fig2} we see another 
interesting and surprising result. Whenever $p_{\!_I}>0.5$ we have a scenario where \textit{more} noise means
\textit{more} efficiency. Indeed, for $p_{\!_I}>0.5$ we can see that 
$\langle\overline{F}_{_{BF,\varnothing,\varnothing}}\rangle$ (solid-black curve) is below the classical limit (dashed curve)
for all values of $p_{\!_B}$ (the quantum teleportation protocol is useless in this situation). 
However, by adding more noise to the protocol, i.e., by putting Bob's qubit in a noisy environment described by the bit
flip map, we can increase the efficiency of the protocol and surpass the $2/3$ limit for
$p_{\!_B}>0.5$. This is illustrated looking at the curve for $\langle\overline{F}_{_{BF,\varnothing,BF}}\rangle$, the 
red-circle curve in Fig. \ref{fig2}.
On the other hand, for $p_{\!_I}<0.5$ the bit flip noise when acting on Bob's qubit always decreases the efficiency of the protocol.

It is worth mentioning that a similar fact occurs when the input qubit is not subjected to noise but the two qubits of the quantum channel are. 
In this scenario Ref. \cite{kno14} shows that when both qubits of the channel are
subjected to the amplitude damping noise we have a better performance than the case where 
only one of the qubits of the channel is acted by this type of noise.

Another interesting scenario happens when noise is unavoidable but Bob can choose the noisy environment in which he keeps his qubit
during the execution of the teleportation protocol. In this case the optimal noise depends in a non-trivial
way on the values of $p_{\!_I}$ and $p_{\!_B}$. For example, when $p_{\!_I}<0.3$ the protocol achieves a better
performance if Bob's qubit is subjected to the amplitude damping noise whenever $p_{\!_B}$ does not exceed 
$\approx 0.6$. However, if $p_{\!_B}$ is greater than $\approx 0.6$ we get a better result if 
Bob's qubit is subjected to the phase flip noise (see the top panels of Fig. \ref{fig2}). 

Let us now move to the case where the input qubit is always subjected to the phase flip noise while Bob's qubit
can suffer any one of the four kinds of noise given in Sec. \ref{noise}. The optimal efficiencies
are now
\begin{eqnarray}
\langle\overline{F}_{_{PhF,\varnothing,PhF}}\rangle &=& \frac{2}{3} \left[1+\frac{|(1-2p_{\!_I})(1-2p_{\!_B})|}{2}\right],
\label{PhFZPhFOpt}\\ 
\langle\overline{F}_{_{PhF,\varnothing,BF}}\rangle&=& \frac{2}{3} \left[1-\frac{p_{\!_B}}{2}+\frac{|1-2p_{\!_I}|(1-p_{\!_B})}{2}\right],
\label{PhFZBFOpt}\\
\langle\overline{F}_{_{PhF,\varnothing,D}}\rangle  &=&  \frac{2}{3} \left[1-\frac{p_{\!_B}}{4}+\frac{|1-2p_{\!_I}|(1-p_{\!_B})}{2}\right],
\label{PhFZDOpt}\\
\langle\overline{F}_{_{PhF,\varnothing,AD}}\rangle &=&  \frac{2}{3} \left[1-\frac{p_{\!_B}}{4}+\frac{p_{\!_B}\cos(2\theta)}{4}\right.
\nonumber \\
&&\left. +\frac{(1-2p_{\!_I})\sqrt{1-p_{\!_B}}\sin(2\theta)}{2}\right].
\label{PhFZADOpt}
\end{eqnarray}
The optimal parameters leading to Eq.~(\ref{PhFZPhFOpt}) are $\theta=\varphi=\pm\pi/4$ if $(1-2p_{\!_I})(1-2p_{\!_B})>0$
and $\theta=-\varphi=\pm\pi/4$ if $(1-2p_{\!_I})(1-2p_{\!_B})<0$. In Eqs.~(\ref{PhFZBFOpt}) and (\ref{PhFZDOpt}) we have
$\theta=\varphi=\pm\pi/4$ if $(1-2p_{\!_I})>0$ and $\theta=-\varphi=\pm\pi/4$ if $(1-2p_{\!_I})<0$.
In Eq.~(\ref{PhFZADOpt}) a possible set of optimal parameters is such that $\varphi=\pi/4$  and 
$\theta$ given by the solution to $d\langle\overline{F}_{_{PhF,\varnothing,AD}}\rangle/d\theta$ $=$ $0$, i.e,
\begin{equation}
\tan(2\theta)=\frac{2(1-2p_{\!_I})\sqrt{1-p_{\!_B}}}{p_{\!_B}},
\label{opttheta2}
\end{equation}
where $\cos(2\theta)>0$, $\sin(2\theta)>0$ if $p_{\!_I}<1/2$, and $\sin(2\theta)<0$ if $p_{\!_I}>1/2$. 
Note that Eq.~(\ref{opttheta2}) implies that the greatest efficiency
is achieved with less entanglement whenever $p_{\!_B}\neq 0$.

\begin{figure}[!ht]
\includegraphics[width=8cm]{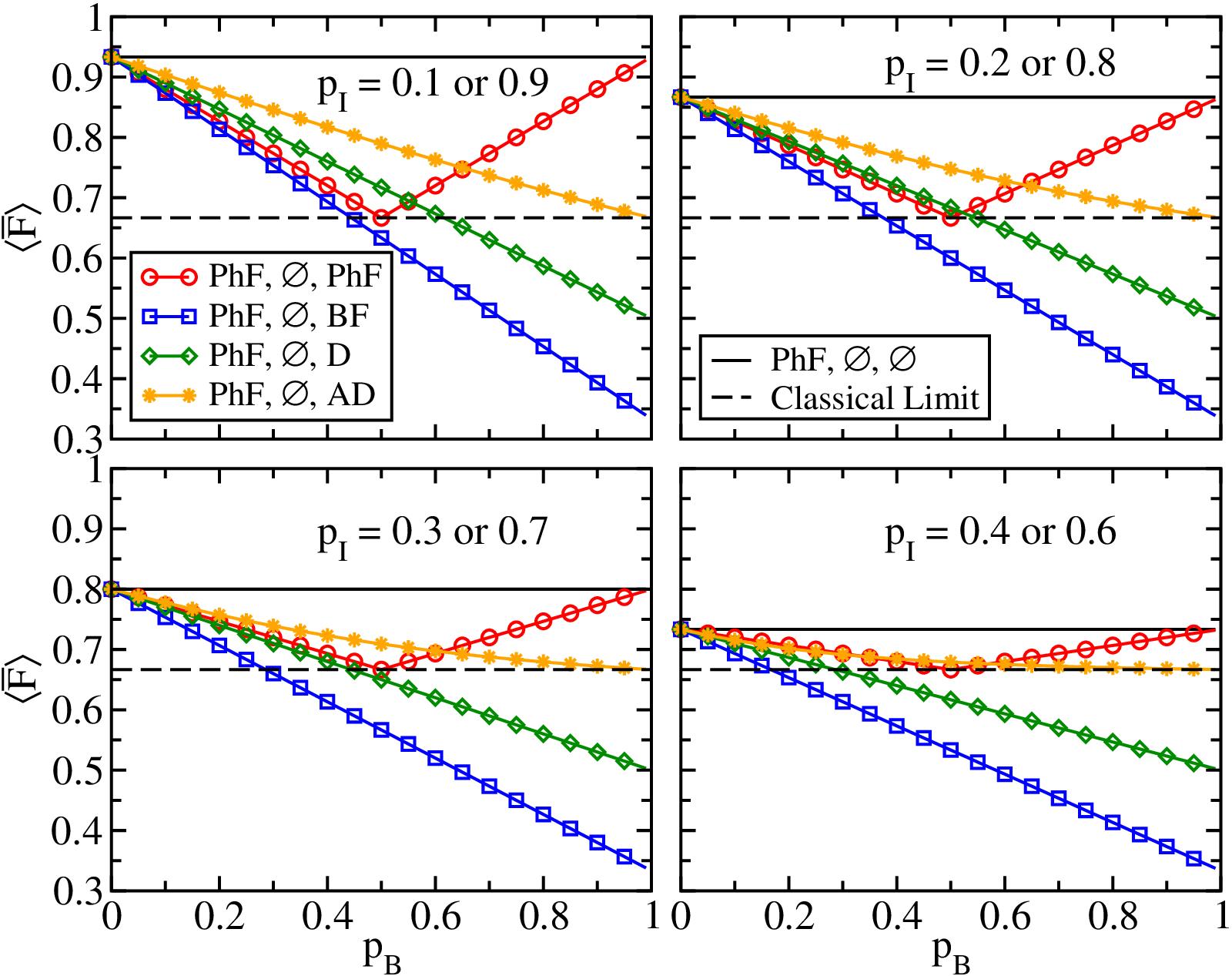}
\caption{\label{fig3}(color online) Efficiency of the teleportation protocol when both the  
input qubit ($p_{\!_I}$) and Bob's qubit ($p_{\!_B}$) are affected by a noisy environment.  
The dashed line, $\langle\overline{F}\rangle=2/3$, marks the value below which classical protocols
(no entanglement) give the same efficiency \cite{bra00}. Here the input qubit is always subjected
to the phase flip ($PhF$) noise while Bob's qubit may suffer from several types of noise.}
\end{figure}

In Fig. \ref{fig3} we plot Eqs.~(\ref{PhFZPhFOpt}) to (\ref{PhFZADOpt}) as a function of $p_{\!_B}$
for several values of $p_{\!_I}$. 
Now, contrary to the case where the input qubit is subjected to the bit flip noise, 
the addition of more noise to the protocol by putting Bob's qubit in a noisy environment ($p_{\!_B}\neq 0$)  
does not improve its efficiency when compared to the noiseless case ($p_{\!_B}=0$).  However,
if noise is inevitable and Bob can choose among different noise channels, he can improve the efficiency
of the protocol by properly selecting the right noise. His choice depends on the probability $p_{\!_B}$ 
for the noise to act on his qubit or, equivalently, on the time his qubit is subjected to a particular 
noisy environment. This is illustrated in Fig. \ref{fig3}, where we note that for $p_{\!_B}$ ranging 
from zero to $\approx 0.6$, the best performance is achieved if Bob's qubit is subjected to the amplitude damping noise (orange-star curve). 
On the other hand, from $p_{\!_B}\approx 0.6$ to one, 
it is better to have both the input and Bob's qubit 
subjected to same type of noise (red-circle curve).

We have also investigated the two remaining cases, i.e., when the input qubit is subjected either to
the depolarizing or to the amplitude damping noise. The qualitative
behavior for the efficiency of the teleportation protocol are similar to the phase flip noise
just described and specific quantitative details are given in the Appendix. 

\subsection{Noise in all Alice's qubits}

We now want to investigate the scenario where all qubits with Alice 
are subjected to the same type of noise. This scenario is relevant,
for example, when the teleportation protocol is employed for quantum communication tasks and
it is Alice that generates the input and the entangled channel. 
In this case the input qubit and Alice's share of the entangled state always lie
in the same environment and are thus subjected to the same type of noise during the same time span.
This latter fact means that $p_{\!_I}=p_{\!_A}=p$. 
Bob's qubit, on the other hand, travels from Alice to Bob and may suffer a different
type of noise. 

If Alice's qubits are subjected to the bit flip noise we have the following 
optimal efficiencies according to the noise suffered by Bob's qubit,
\begin{eqnarray}
\langle\overline{F}_{_{BF, BF, \varnothing}}\rangle &=& 1-\frac{4p(1-p)}{3},
\label{BFBFZOpt}\\ 
\langle\overline{F}_{_{BF, BF, BF}}\rangle&=& 1-\frac{2p_{\!_B}}{3}-\frac{4p(1-p)(1-2p_{\!_B})}{3},
\label{BFBFBFOpt}\\
\langle\overline{F}_{_{BF,BF, PhF}}\rangle  &=&\frac{2}{3}-\frac{2p(1-p)}{3} \nonumber \\
&&+\frac{[1-2p(1-p)]|1-2p_{\!_B}|}{3},
\label{BFBFPhFOpt}\\
\langle\overline{F}_{_{BF, BF, D}}\rangle &=& 1-\frac{p_{\!_B}}{2}-\frac{4p(1-p)(1-p_{\!_B})}{3},
\label{BFBFDOpt}\\
\langle\overline{F}_{_{BF, BF, AD}}\rangle &=& \frac{2}{3}-\frac{p_{\!_B}}{6}-\frac{2p(1-p)(1-p_{\!_B})}{3}
\nonumber \\
&+&\frac{(1-2p)^2p_{\!_B}\cos(2\theta)}{6}
\nonumber \\
&+&\frac{[1-2p(1-p)]\sqrt{1-p_{\!_B}}\sin(2\theta)}{3}.
\label{BFBFADOpt}
\end{eqnarray}
The optimal parameters giving Eqs.~(\ref{BFBFZOpt}), (\ref{BFBFBFOpt}), and (\ref{BFBFDOpt}) are $\theta=\varphi=\pm\pi/4$ while
for Eq.~(\ref{BFBFPhFOpt}) we have $\theta=\varphi=\pm\pi/4$ if $p_{\!_B}<1/2$ and $\theta=-\varphi=\pm\pi/4$ if $p_{\!_B}>1/2$.
In Eq.~(\ref{BFBFADOpt}) a possible set of optimal parameters are $\varphi=\pi/4$ with $\theta$ given by 
\begin{equation}
\tan(2\theta)=\frac{2[1-2p(1-p)]\sqrt{1-p_{\!_B}}}{(1-2p)^2p_{\!_B}},
\end{equation}
where $\cos(2\theta)>0$ and $\sin(2\theta)>0$.

\begin{figure}[!ht]
\includegraphics[width=8cm]{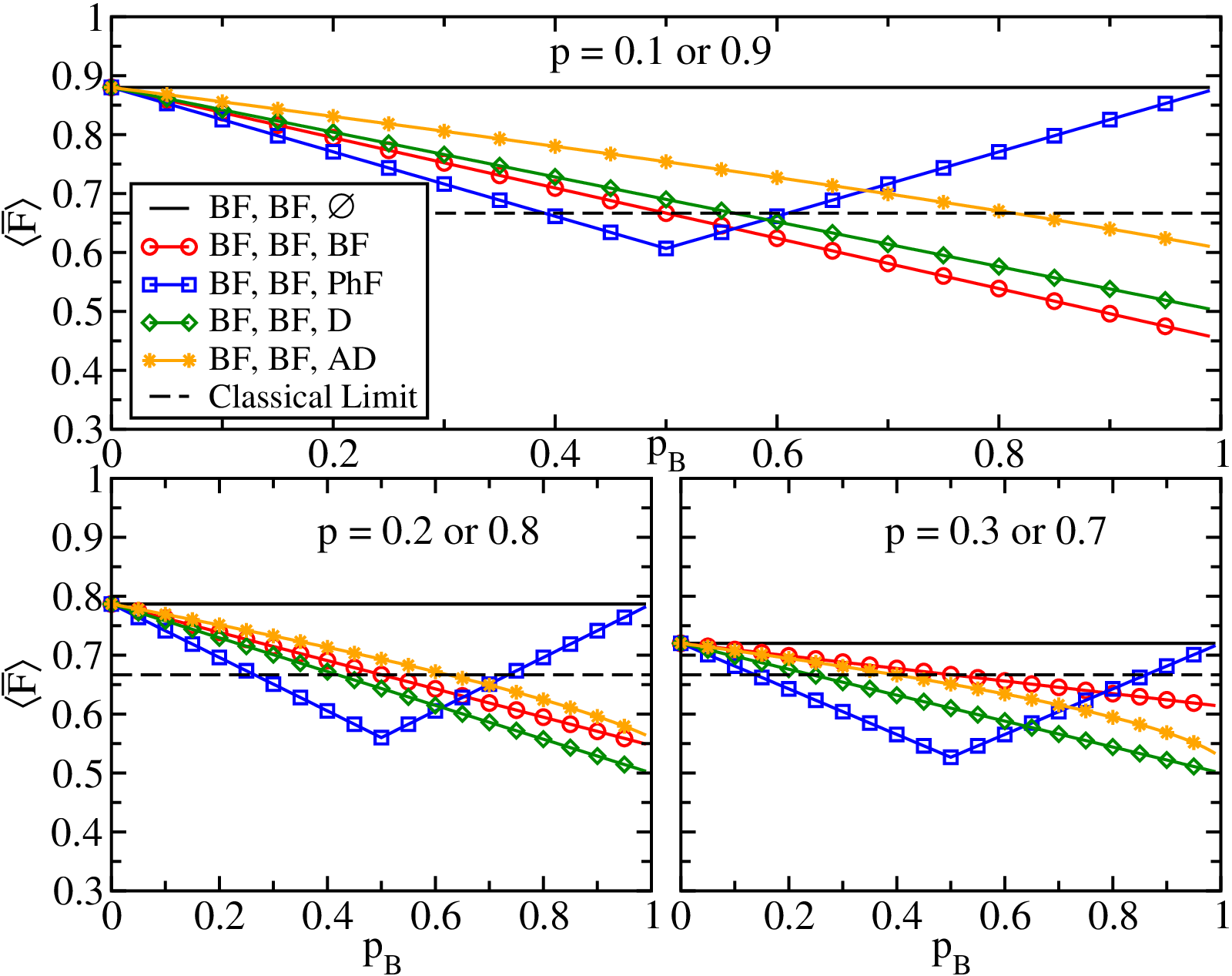}
\caption{\label{fig4}(color online) Efficiency of the teleportation protocol when Alice's qubits suffer the
bit flip noise ($p_{\!_I}=p_{\!_A}=p$) and Bob's qubit ($p_{\!_B}$) are affected by several types of 
noise. The dashed line, $\langle\overline{F}\rangle=2/3$, marks the value below which classical protocols
(no entanglement) give the same efficiency \cite{bra00}. }
\end{figure}

In Fig. \ref{fig4} we plot Eqs.~(\ref{BFBFZOpt}) to (\ref{BFBFADOpt}) as a function of $p_{\!_B}$
for several values of $p$. 
The first thing we note is that the addition of more noise does not
improve the efficiency of the protocol (no curve crosses the solid-black one).
This is in contrast to the case where only the input qubit is affected by the bit flip noise
and we put Bob's qubit in a noisy environment too (Fig. \ref{fig2}).  
Second, if noise is unavoidable Bob can improve the efficiency
of the protocol by selecting the right noise channel according to the values of $p$
and $p_{\!_B}$. 

We also studied the cases where Alice's qubits are affected by the other
three types of noise and we found that the qualitative behavior of all those cases
are similar to what is shown in Fig. \ref{fig4}. Indeed, we did not find any situation
in which more noise (allowing Bob's qubit to be acted by noise) improves the overall efficiency. 
However, 
for all three cases we noted that whenever Bob's noise channel differs from Alice's we 
may have significant improvement in the efficiency of the protocol. In particular, 
if Bob's qubit is subjected to the amplitude damping noise we achieve the best performance
for low values of $p_{\!_B}$. For high values of $p_{\!_B}$, on the other hand,
the phase flip channel is the one furnishing the best performance.
Finally, when the amplitude damping channel acts on at least one qubit we have observed the same 
trend described before concerning the degree of entanglement 
of the optimal quantum channel, namely, less entanglement means 
better performance.

\subsection{Noise in the quantum channel}

Another important case is the one where the channel qubits 
are subjected to the same type of noise. This may occur in
the implementation of quantum communication protocols in which 
the quantum channel is generated by a third
party symmetrically positioned between Alice and Bob. In the case where both qubits
find similar noisy environments during their trip to Alice and Bob, 
they will be acted by the noise during the same amount of time, which implies
$p_{\!_A}=p_{\!_B}=p$. 
The input qubit, on the other hand, may suffer a different
type of noise. 

If the channel qubits are subjected to the bit flip noise we have the following 
optimal efficiencies depending on the type of noise acting on the input qubit,
\begin{eqnarray}
\langle\overline{F}_{_{\varnothing, BF, BF}}\rangle &=& 1-\frac{4p(1-p)}{3},
\label{ZBFBFOpt}\\ 
\langle\overline{F}_{_{BF, BF, BF}}\rangle&=& 1 - \frac{2p_{\!_I}}{3} - \frac{4(1-2p_{\!_I})p(1-p)}{3},
\label{BFBFBFOpt2}\\
\langle\overline{F}_{_{PhF, BF, BF}}\rangle  &=&\frac{2}{3}-\frac{2p(1-p)}{3} \nonumber \\
&&+\frac{|1-2p_{\!_I}|[1-2p(1-p)]}{3} ,
\label{PhFBFBFOpt}\\
\langle\overline{F}_{_{D, BF, BF}}\rangle &=& 1-\frac{p_{\!_I}}{2}-\frac{4(1-p_{\!_I})p(1-p)}{3},
\label{DBFBFOpt}\\
\langle\overline{F}_{_{AD, BF, BF}}\rangle &=& \frac{2}{3}-\frac{p_{\!_I}}{6}-\frac{2(1-p_{\!_I})p(1-p)}{3}
\nonumber \\
&+&\frac{\sqrt{1-p_{\!_I}}[1-2p(1-p)]}{3}.
\label{ADBFBFOpt}
\end{eqnarray}
The optimal parameters giving Eqs.~(\ref{ZBFBFOpt}), (\ref{BFBFBFOpt2}), (\ref{DBFBFOpt}), and (\ref{ADBFBFOpt}) are $\theta=\varphi=\pm\pi/4$ and 
for Eq.~(\ref{PhFBFBFOpt}) we have $\theta=\varphi=\pm\pi/4$ if $p_{\!_I}<1/2$ and $\theta=-\varphi=\pm\pi/4$ if $p_{\!_I}>1/2$. 

Comparing Eqs.~(\ref{BFBFZOpt})-(\ref{BFBFADOpt}) with Eqs.~(\ref{ZBFBFOpt})-(\ref{ADBFBFOpt}) we see that
$\langle\overline{F}_{_{BF, BF, X}}\rangle=\langle\overline{F}_{_{X, BF, BF}}\rangle$, $X = \varnothing, BF, PhF, D, AD$, if $p_{\!_B} = p_{\!_I}$ 
and $\theta=\pm \pi/4$ in Eq.~(\ref{BFBFADOpt}). This means that the same analysis and discussions given in the previous section apply here.
The only quantitative difference occurs for Eq.~(\ref{ADBFBFOpt}) since now, contrary to (\ref{BFBFADOpt}), the optimal channel is a 
maximally entangled state. However, the qualitative behavior for the efficiency given by Eq.~(\ref{BFBFADOpt}) applies here too, in particular
the fact that it is the combination of noise channels leading to the best performance for low $p_{\!_I}$ whenever $p$ is no greater than $\approx 0.3$. 

A direct calculation also shows that $\langle\overline{F}_{_{PhF, PhF, X}}\rangle=\langle\overline{F}_{_{X, PhF, PhF}}\rangle$
and $\langle\overline{F}_{_{D, D, X}}\rangle=\langle\overline{F}_{_{X, D, D}}\rangle$, 
$X$ $=$ $\varnothing$, $BF$, $PhF$, $D$, $AD$, if $p_{\!_B} = p_{\!_I}$ and $\theta=\pm \pi/4$ in the expressions for
$\langle\overline{F}_{_{PhF, PhF, AD}}\rangle$ and $\langle\overline{F}_{_{D, D, AD}}\rangle$. On the other hand, this symmetry does
not hold for $\langle\overline{F}_{_{X, AD, AD}}\rangle$. However, its qualitative behavior still
resembles the one for $\langle\overline{F}_{_{AD, AD, X}}\rangle$, in the sense that 
$\langle\overline{F}_{_{AD, AD, AD}}\rangle$ gives the best performance for low $p_{\!_I}$ while for
high $p_{\!_I}$ the best scenario is $\langle\overline{F}_{_{PhF, AD, AD}}\rangle$.

\section{A different quantum channel}
\label{difchannel}

Another aspect affecting the efficiency of the teleportation protocol in a noisy environment is 
related to the choice of the quantum channel, even if we choose among those having the same amount of entanglement \cite{ban02}. 
Throughout this article we have employed the generalized Bell state $|B^\theta_1\rangle$, which reduces to
the maximally entangled Bell state $|\Phi^+\rangle$ when $\theta = \pi/4$ (see Eq.~(\ref{B1})). However, if we use the state
$|B^\theta_3\rangle$, which approaches the maximally entangled Bell state $|\Psi^+\rangle$ as $\theta \rightarrow \pi/4$
(see Eq.~(\ref{B3})), we may get a different efficiency.

\begin{figure}[!ht]
\includegraphics[width=8cm]{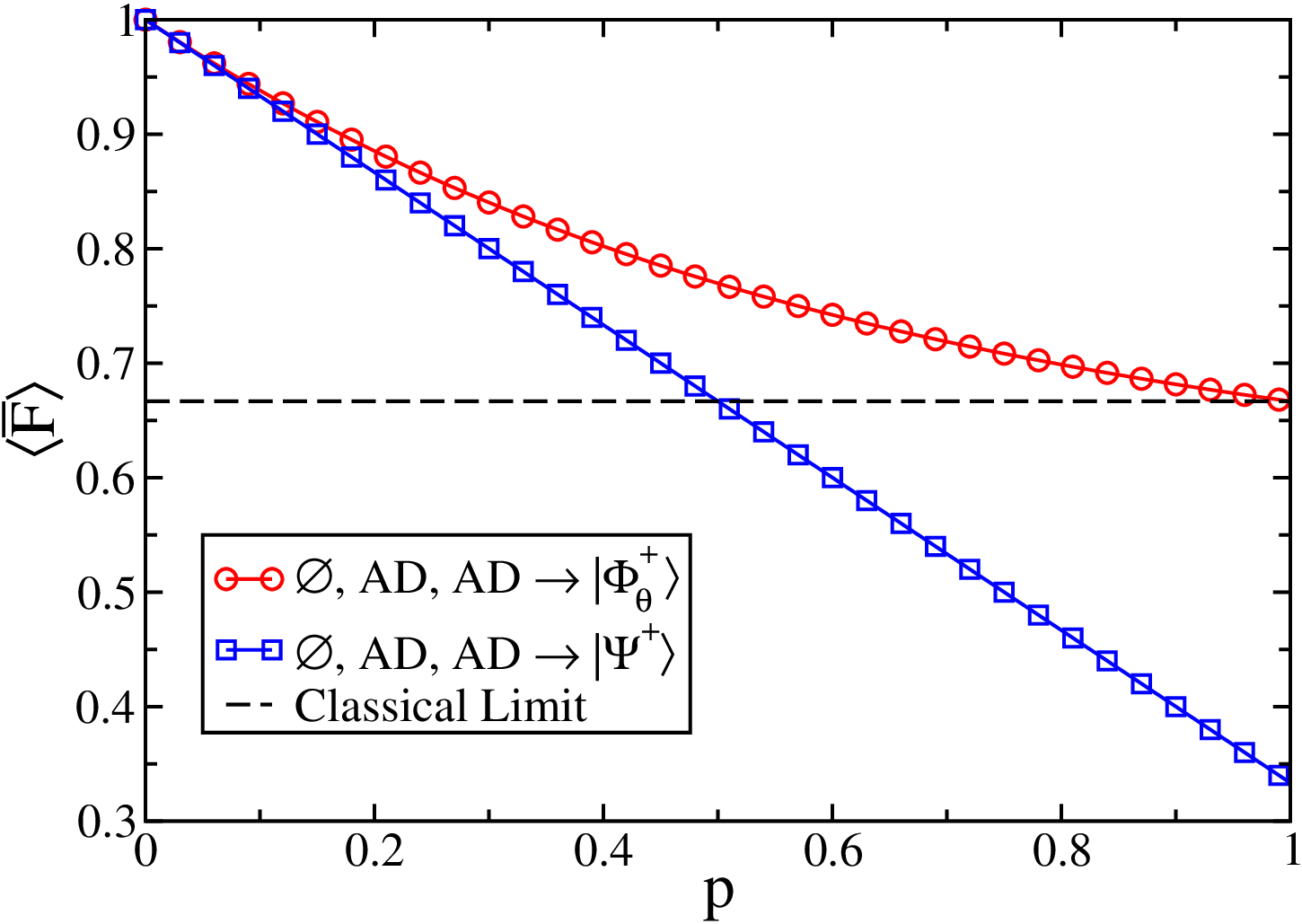}
\caption{\label{fig5}(color online) Efficiency of the teleportation protocol for different channels  
subjected to the same noise, namely, the amplitude damping noise ($p_{\!_A}=p_{\!_B}=p$), with the input qubit 
in a noiseless environment. The dashed line, $\langle\overline{F}\rangle=2/3$, marks the value below which classical protocols
(no entanglement) give the same efficiency \cite{bra00}.}
\end{figure}

For example, studying the case where both qubits of the quantum channel are acted by the same type of noise during the same time
($p_{\!_A}=p_{\!_B}=p$), and employing either the quantum channel $|B^\theta_1\rangle$ or $|B^\theta_3\rangle$, we obtain
$
\langle\overline{F}_{_{\varnothing, X, X}}\rangle_{\!_{|B^\theta_1\rangle}}
=
\langle\overline{F}_{_{\varnothing, X, X}}\rangle_{\!_{|B^\theta_3\rangle}}
$
for all $X=BF, PhF, D$ but a different efficiency when $X=AD$. This last fact is illustrated in 
Eqs.~(\ref{ZADADPhiOpt}) and (\ref{ZADADPsiOpt}),
\begin{eqnarray}
\langle\overline{F}_{_{\varnothing, AD, AD}}\rangle_{\!_{|\Phi^+_\theta\rangle}} &=& \frac{2}{3}+\frac{(1-p)[\sqrt{1+p^2}-p]}{3},
\label{ZADADPhiOpt}\\
\langle\overline{F}_{_{\varnothing, AD, AD}}\rangle_{\!_{|\Psi^+\rangle}}  &=& 1-\frac{2p}{3},
\label{ZADADPsiOpt}
\end{eqnarray}
where the optimal parameters giving Eq.~(\ref{ZADADPhiOpt}) are $\varphi=\pi/4$ and $\tan(2\theta)=1/p$, with
$0\leq \theta \leq \pi/2$, and the optimal ones leading to Eq.~(\ref{ZADADPsiOpt}) are such that $\theta=\varphi=\pm\pi/4$.
It is not difficult to see that 
$
\langle\overline{F}_{_{\varnothing, AD, AD}}\rangle_{\!_{|B^\theta_1\rangle}}
>
\langle\overline{F}_{_{\varnothing, AD, AD}}\rangle_{\!_{|B^\theta_3\rangle}}
$
for  $0<p\leq 1$. In Fig. \ref{fig5} we plot 
Eqs.~(\ref{ZADADPhiOpt}) and (\ref{ZADADPsiOpt}) as a function of $p$. 

We can qualitatively understand why we have different efficiencies using different Bell states if we note, first, that 
$|B^\theta_1\rangle$ is a superposition of $|00\rangle$ and $|11\rangle$ while $|B^\theta_3\rangle$
is given by the superposition of $|01\rangle$ and $|10\rangle$ and, second, that the
amplitude damping channel models energy dissipation and thus does not affect the ground state $|0\rangle$. With that in mind,
we see that both states that make $|B^\theta_3\rangle$, $|01\rangle$ and $|10\rangle$, are affected by this type of noise
when it acts on both qubits. For $|B^\theta_1\rangle$, however, only $|11\rangle$ is affected, which makes it more robust for this type
of noise.

\section{Conclusion}

We studied how the efficiency of the quantum teleportation protocol 
is affected when the qubits employed in its execution lie
in a noisy environment. In order to model the noisy environment we employed the most common 
noise channels that one encounters in any realistic implementation of quantum communication protocols,
namely, the bit flip, phase flip (phase damping), depolarizing, and amplitude damping channels. 
We also studied many noise scenarios where one, two or all three qubits
required for the implementation of the teleportation protocol are subjected to noise. 

The first remarkable result we obtained is related to the entanglement needed to get 
the optimal efficiency when noise is present in both the input qubit and in the quantum channel. 
Specifically, we observed a non-trivial interplay
among which quantum channel is used, its amount of entanglement, and the type of noise acting on
it. Indeed, for certain types of quantum channels acted by the amplitude damping noise, we showed that 
\textit{less} entanglement means \textit{more} efficiency, a similar behavior observed in
Ref. \cite{ban12} when noise acts only on the channel qubits. For other channels with the 
same initial entanglement we showed that this behavior is not seen. 
The results given here and in Ref. \cite{ban12}
are a counterintuitive fact since for pure state inputs and pure state channels (no noise) 
more entanglement leads to more efficiency.

Second, we showed a scenario where \textit{more} noise leads to \textit{more} efficiency.
This fact occurred when the input qubit, the one to be teleported from Alice to Bob, lies
in a noisy environment described by the bit flip noise and Bob's qubit, his share of the 
entangled channel, is also affected by this same noise. The efficiency in this case is 
considerably greater when compared to the situation where only the input qubit is subjected to 
this type of noise. This kind of behavior was also observed in Ref. \cite{kno14} when the
channel qubits are subjected to the amplitude damping noise.

Third, when noise is unavoidable but either Alice or Bob can choose the kind of noise that  
acts on their qubits, we showed that the optimal combination of noisy environments leading to
the highest efficiencies is related in a non-trivial way to the time (or probability) the qubits
are affected by the noise. In many scenarios we showed that Alice and Bob should keep their qubits in
different noisy environments  in order to get the best performance for the teleportation protocol.
Furthermore, we showed that sometimes the optimal efficiency is obtained by letting 
one of the qubits be subjected for a longer time to noise than the other one.

Fourth, in a noisy environment we showed that the choice of the quantum channel can affect the efficiency of the teleportation protocol, 
even if we deal with channels having the same amount of entanglement \cite{ban02}. 
Being more specific, we showed that if the quantum channel is a Bell state and 
the amplitude damping noise acts on both qubits of the channel, different Bell states 
lead to different performances for the teleportation protocol.

In summary, the main message we can draw from the results given in this paper, and from the complementary ones
given in Refs. \cite{ban02,ban12,kno14}, is quite clear: 
When noise is taken into account in analyzing the performance of a quantum communication 
protocol we should not expect the optimal settings for the noiseless case to hold anymore.  
Moreover, sometimes counterintuitive optimal settings occur and, therefore, we should always analyze the 
types of noise we are going to face in any realistic implementation of a particular protocol 
and determine those optimal settings on a case-by-case basis.

\begin{acknowledgments}
RF thanks CAPES (Brazilian Agency for the Improvement of Personnel of Higher Education)
for funding and GR thanks the Brazilian agencies CNPq
(National Council for Scientific and Technological Development) and
CNPq/FAPESP (State of S\~ao Paulo Research Foundation) for financial support through the National Institute of
Science and Technology for Quantum Information.
\end{acknowledgments}

\appendix*

\section{}

The optimal efficiencies assuming the input qubit is subjected to the depolarizing noise and
Bob's qubit to the four different types of noise are
\begin{eqnarray}
\langle\overline{F}_{_{D,\varnothing,D}}\rangle &=& 1 - \frac{p_{\!_I}}{2}-\frac{p_{\!_B}(1-p_{\!_I})}{2},
\label{DZDOpt}\\ 
\langle\overline{F}_{_{D,\varnothing,BF}}\rangle&=& 1 - \frac{p_{\!_I}}{2}-\frac{2p_{\!_B}(1-p_{\!_I})}{3},
\label{DZBFOpt}\\
\langle\overline{F}_{_{D,\varnothing,PhF}}\rangle  &=& \frac{2}{3} - \frac{p_{\!_I}}{6}+\frac{(1-p_{\!_I})|1-2p_{\!_B}|}{3},
\label{DZPhFOpt}\\
\langle\overline{F}_{_{D,\varnothing,AD}}\rangle &=&  \frac{2}{3} - \frac{p_{\!_I}}{6}-\frac{(1-p_{\!_I})p_{\!_B}[1-\cos(2\theta)]}{6}
\nonumber \\
&& + \frac{(1-p_{\!_I})\sqrt{1-p_{\!_B}}\sin(2\theta)}{3}.
\label{DZADOpt}
\end{eqnarray}
The optimal settings for Eqs.~(\ref{DZDOpt}) and (\ref{DZBFOpt}) are $\theta=\varphi=\pm \pi/4$,
for Eq.~(\ref{DZPhFOpt}) are $\theta=\varphi=\pm \pi/4$ if $p_{\!_B} < 1/2$ and
$\theta=-\varphi=\pm \pi/4$ if $p_{\!_B}>1/2$, and for Eq.~(\ref{DZADOpt}) we have
$\varphi=\pi/4$ and 
\begin{equation}
\tan(2\theta) = \frac{2\sqrt{1-p_{\!B}}}{p_{\!_B}},
\end{equation}
such that $\sin(2\theta)>0$ and $\cos(2\theta)>0$.

\begin{figure}[!ht]
\includegraphics[width=8cm]{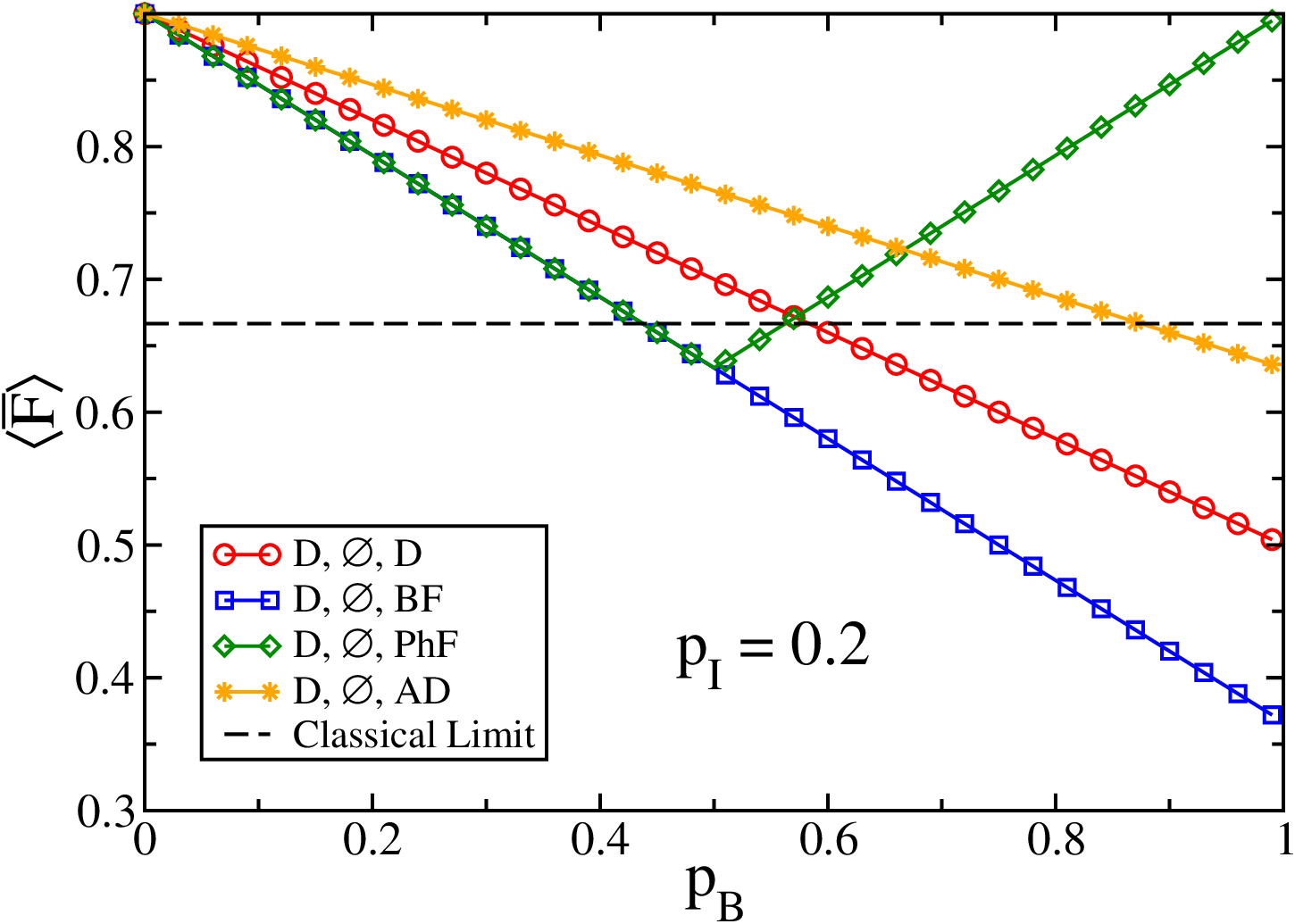}
\caption{\label{figA1}(color online) Efficiency of the teleportation protocol when both the  
input qubit ($p_{\!_I}$) and Bob's qubit ($p_{\!_B}$) are affected by a noisy environment.  
The dashed line, $\langle\overline{F}\rangle=2/3$, marks the value below which classical protocols
(no entanglement) give the same efficiency \cite{bra00}. Here the input qubit is always subjected
to the depolarizing ($D$) noise while Bob's qubit may suffer from several types of noise.}
\end{figure}

In Fig. \ref{figA1} we illustrate the behavior of Eqs.~(\ref{DZDOpt}) to (\ref{DZADOpt}) for a 
particular value of $p_{\!_I}$. It is worth mentioning that as we increase the value of $p_{\!_I}$ we obtain
the same curves with all points translated to lower values of $\langle \overline{F}\rangle$. For values of
$p_{\!_I}$ greater than $\approx 0.7$ we do not have any curve above the classical $2/3$ limit. 

\begin{figure}[!ht]
\includegraphics[width=8cm]{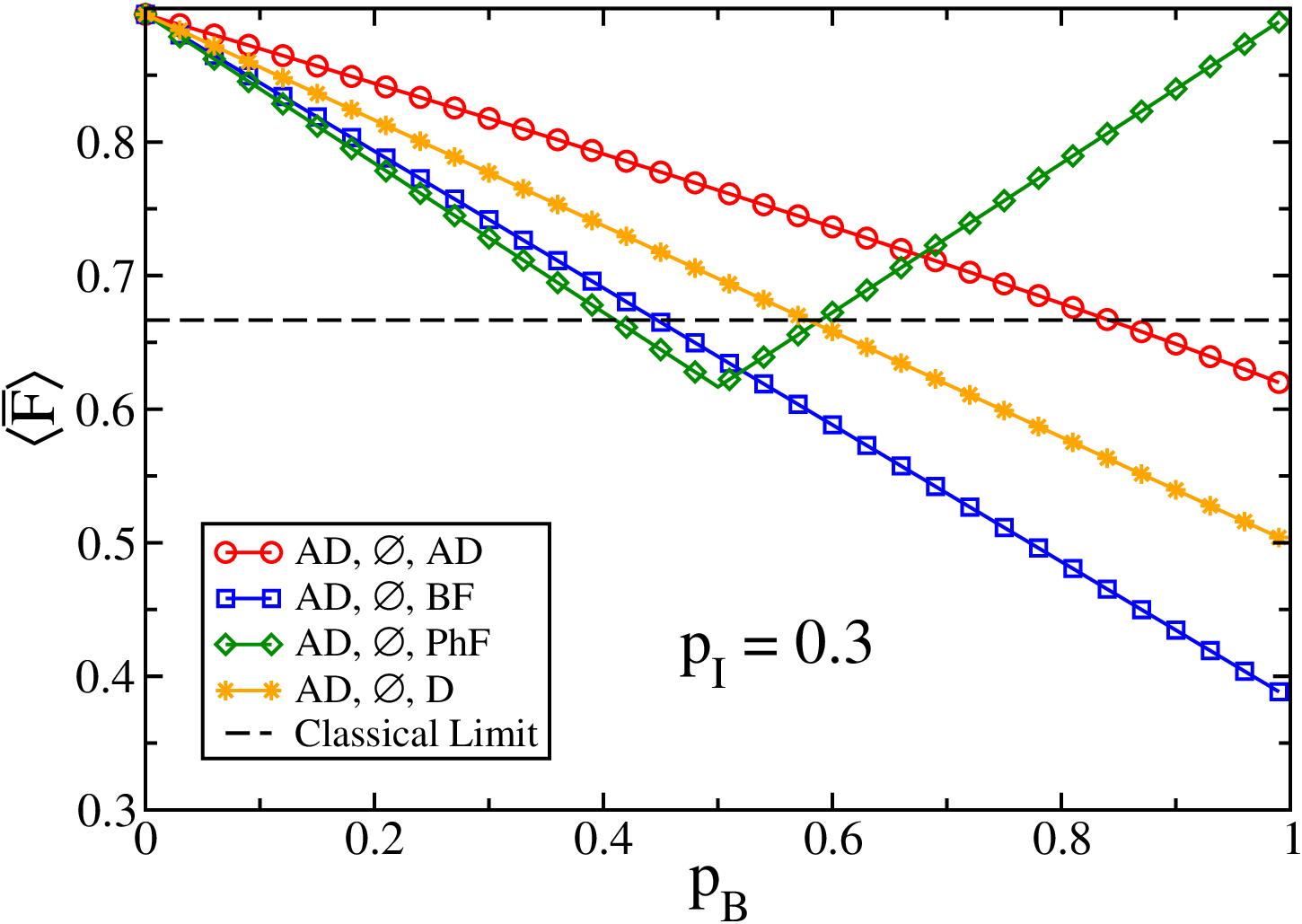}
\caption{\label{figA2}(color online) Efficiency of the teleportation protocol when both the  
input qubit ($p_{\!_I}$) and Bob's qubit ($p_{\!_B}$) are affected by a noisy environment.  
The dashed line, $\langle\overline{F}\rangle=2/3$, marks the value below which classical protocols
(no entanglement) give the same efficiency \cite{bra00}. Here the input qubit is always subjected
to the amplitude damping ($AD$) noise while Bob's qubit may suffer from several types of noise.}
\end{figure}

Finally, the optimal efficiencies when the input qubit is subjected to the amplitude damping noise are
\begin{eqnarray}
\langle\overline{F}_{_{AD,\varnothing,AD}}\rangle &=& \frac{2}{3} - \frac{p_{\!_I}}{6}-\frac{(1-p_{\!_I})p_{\!_B}[1-\cos(2\theta)]}{6}
\nonumber \\
&& + \frac{\sqrt{(1-p_{\!_I})(1-p_{\!_B})}\sin(2\theta)}{3},
\label{ADZADOpt}\\ 
\langle\overline{F}_{_{AD,\varnothing,BF}}\rangle&=& \frac{2}{3} - \frac{p_{\!_I}}{6}-\frac{(1-p_{\!_I})p_{\!_B}}{3}
\nonumber \\
&& + \frac{\sqrt{1-p_{\!_I}}(1-p_{\!_B})}{3},
\label{ADZBFOpt}\\
\langle\overline{F}_{_{AD,\varnothing,PhF}}\rangle  &=&  \frac{2}{3} - \frac{p_{\!_I}}{6}+\frac{\sqrt{1-p_{\!_I}}|1-2p_{\!_B}|}{3},
\label{ADZPhFOpt}\\
\langle\overline{F}_{_{AD,\varnothing,D}}\rangle &=&  \frac{2}{3} - \frac{p_{\!_B}}{6} + 
\frac{1-p_{\!_B}}{3}\hspace{-.1cm}\left(\hspace{-.1cm} \sqrt{1\!-\!p_{\!_I}} \!-\!\frac{p_{\!_I}}{2}\hspace{-.1cm} \right)\hspace{-.1cm}.
\label{ADZDOpt}
\end{eqnarray}
The optimal settings for Eq.~(\ref{ADZADOpt}) are $\varphi=\pi/4$ and
\begin{equation}
\tan(2\theta) = \frac{2}{p_{\!_B}}\sqrt{\frac{1-p_{\!B}}{1-p_{\!I}}},
\end{equation}
such that $\sin(2\theta)>0$ and $\cos(2\theta)>0$,
for Eqs. (\ref{ADZBFOpt}) and (\ref{ADZDOpt}) are $\theta=\varphi=\pm \pi/4$,
and for Eq.~(\ref{ADZPhFOpt}) are $\theta=\varphi=\pm \pi/4$ if $p_{\!_B} < 1/2$ and
$\theta=-\varphi=\pm \pi/4$ if $p_{\!_B}>1/2$.

In Fig. \ref{figA2} we plot Eqs.~(\ref{ADZADOpt}) to (\ref{ADZDOpt}) for a 
specific value of $p_{\!_I}$. For other values of $p_{\!_I}$ we have similar curves and similar relations among
the different curves. For $p_{\!_I}$ greater than $\approx 0.9$ we do not have any curve crossing the classical limit.


\begin{thebibliography}{200}
\bibitem{note1} In order to save space, we abuse notation and use the same
symbol $P_j^\varphi$ to denote $P_j^\varphi\otimes \mathbb{1}_B$, where
$\mathbb{1}_B$ is the identity matrix acting on Bob's qubit. 
$P_j^\varphi\otimes \mathbb{1}_B$ is the correct mathematical representation of
Alice's projective operator when acting on a three qubit state.
\end{thebibliography}
\end{document}